\documentclass[conference]{IEEEtran}
\usepackage[colorlinks,urlcolor=blue,linkcolor=blue,citecolor=blue]{hyperref}
\usepackage{cite}
\usepackage{amsmath,amssymb,amsfonts}
\usepackage{graphicx}
\usepackage{textcomp}
\usepackage{xcolor}
\usepackage{amsmath}
\usepackage{float} 
\usepackage{booktabs}
\usepackage[ruled,linesnumbered]{algorithm2e}
\usepackage{algpseudocode}
\usepackage[normalem]{ulem}
\usepackage{subcaption}

\def\BibTeX{{\rm B\kern-.05em{\sc i\kern-.025em b}\kern-.08em
    T\kern-.1667em\lower.7ex\hbox{E}\kern-.125emX}}

\begin{document}

\title{Connectivity-Aware Pheromone Mobility Model for Autonomous UAV Networks}

\author{
\IEEEauthorblockN{
Shreyas~Devaraju\IEEEauthorrefmark{1}, Alexander~Ihler\IEEEauthorrefmark{2}, and Sunil~Kumar
\IEEEauthorrefmark{1}\IEEEauthorrefmark{3}
}
\IEEEauthorblockA{
\IEEEauthorrefmark{1}Computational Science Research Center, San Diego State University, San Diego, CA, USA\\
\IEEEauthorrefmark{2}School of Information \& Computer Science, University of California, Irvine, CA, USA\\
\IEEEauthorrefmark{3}Electrical \& Computer Engineering Department, San Diego State University, San Diego, CA, USA\\
Email: sdevaraju@sdsu.edu, ihler@ics.uci.edu, skumar@sdsu.edu
}}

\maketitle

\begin{abstract}
UAV networks consisting of reduced size, weight, and power (low SWaP) fixed-wing UAVs are used for civilian and military applications such as search and rescue, surveillance, and tracking. To carry out these operations efficiently, there is a need to develop scalable, decentralized autonomous UAV network architectures with high network connectivity. However, the area coverage and the network connectivity requirements exhibit a fundamental trade-off. In this paper, a connectivity-aware pheromone mobility (CAP) model is designed for search and rescue operations, which is capable of maintaining connectivity among UAVs in the network.  We use stigmergy-based digital pheromone maps along with distance-based local connectivity information to autonomously coordinate the UAV movements, in order to improve its map coverage efficiency while maintaining high network connectivity.
\end{abstract}


\section{Introduction}

The unmanned aerial vehicles (UAVs), equipped with self-localization and sensing capabilities, are used in applications such as search-and-rescue, tracking and surveillance \cite{b9,b10,c20,b2}. Distributed UAV networks are scalable, sense simultaneously in an expanded area, and do not have a single point of failure. In distributed or decentralized, autonomous UAV network architectures, the nodes perform only local sensing and communicate with their neighbors without any global knowledge \cite{b9}. However, such networks can face communication issues, since low SWaP (size, weight and power) UAVs have a limited communication range. Therefore, the connectivity among the UAVs must be maintained to allow their coordination and control. Whereas a high network connectivity facilitates better communication among the UAVs, an increase in coverage performance leads to a faster discovery and better tracking of targets in a search area. Note that the area  coverage and network connectivity requirements exhibit a trade-off, i.e., dispersing the UAVs to improve coverage will typically negatively impact connectivity \cite{b10,c20}.

Swarm intelligence methodologies inspired by nature, such as the social behavior of insects, birds, and fish, can be used to solve complex problems cooperatively by using simple rules and local interactions. One such widely used method is the use of stigmergic digital pheromones \cite{b2,b3}, which act as the spatio-temporal potential fields that are used to coordinate and control the UAV movement. 
In this paper, we use the digital pheromone-based stigmergic algorithms to achieve quick exploration of completely unknown environments. Their decentralized nature makes them fault-tolerant and highly scalable. However, most repulsion pheromone-based mobility models focus only on coverage performance of UAV networks, and ignore the connectivity.

This paper addresses the problem of achieving an efficient coverage of a given search area while preserving the network connectivity in an autonomous, decentralized UAV network.
We design a UAV mobility model, which combine the pheromone mobility model with local connectivity information to optimize the coverage and connectivity performance. We call the model as connectivity-aware pheromone mobility (CAP) model. This mobility model selects a UAV path that balances pheromone values with estimated connectivity values at a number of potential waypoints.

Paper Organization: We first review the existing schemes for UAV mobility in Section~\ref{Related_Work_section}, followed by a brief overview of the pheromone based UAV mobility model in Section~\ref{Overview_Pheromone_Mobility_section}. Then we describe our proposed CAP model in Section~\ref{proposed_CAP}. The simulation results are discussed in Section~\ref{SimultionResults}, followed by the conclusions in Section~\ref{conclusion_section}.

\section{Related Work}\label{Related_Work_section}

Several algorithms such as particle swarm optimization, artificial bee colony and ant colony optimization (ACO) have been proposed for control and coordination of swarms for various search, rescue, and tracking applications \cite{b9,pigeon_search}. The digital pheromone based mobility model has been used for target search and other other similar tasks in UAV networks. 

In digital pheromone schemes, information about the pheromone map is communicated between agents in the network through direct or indirect communication. In the direct communication methods, each agent maintains a full or partial pheromone map of its immediate vicinity. Updates in the pheromone map due to deposits or withdrawals are communicated only locally. In \cite{b13}, distributed stigmergic coordination of UAVs for automatic target recognition is done through direct communication. The UAVs mark potential targets and communicate the pheromone information to their neighbors using a decentralized gossip mechanism.

Sauter et al. \cite{b3} is an example of indirect communication scheme for controlling and coordinating the UAV swarm for surveillance, target acquisition, and tracking. Here the coordination of swarm members is based on digital pheromones maintained in an artificial pheromone map, and a centralized base station (BS) is used to communicate the global pheromone map to all the UAVs. Failure of the centralized BS may lead to failure of the entire system. 

Some schemes use a fusion of stigmergic pheromone algorithm and flocking behaviors to coordinate a group of UAVs for performing decentralized target search \cite{b7,b2}. Here, the UAVs deposit digital attract pheromones when a potential target is detected to attract UAVs in the area; Repel pheromones are deposited when no target is found. They also follow Boids \cite{boids} flocking rules to organize the swarm for better perception and communication for tracking the targets. An evolutionary algorithm is used in \cite{b2} for tuning the pheromone and flocking behaviors to get an optimal performance. Shao et al. \cite{b16} designed a navigation algorithm by using the pheromone algorithm on top of the Olfati-Saber’s flocking algorithm \cite{olfati}, where leader-follower based flocking is performed. 
The coverage and network connectivity performance for a UAV group using a random  vs. pheromone guided mobility model are compared in \cite{b10}. While the random model follows a Markov process, the UAVs move to a low repel pheromone area in the pheromone model. The pheromone model provides a better coverage than the random model, but neither model show a good connectivity performance. Messous et al. \cite{c22} address the connectivity issue in UAV fleets by weighting a UAV's tendency to follow its neighbor based on its connectivity, hop count to the base station, and energy level. Similarly, dual-pheromone clustering hybird approach (DPCHA) \cite{c21} uses dual pheromones for target tracking and area coverage, and the clustering to maintain stable network connectivity.

\subsection{Review of CACOC\textsuperscript{2} Model} \label{CACOC2_section}

The  CACOC\textsuperscript{2}  model  \cite{CACOC2}  uses  the  ACO  with  a  chaotic dynamical  system  (CACOC)  \cite{CACOC},  together  with  the  Boids flocking  model  to  maximize  the coverage  while  preserving the network connectivity. The CACOC model \cite{CACOC} uses the pheromone mobility model, along with chaotic dynamics obtained using the Rossler system, to obtain a deterministic but unpredictable system. In CACOC, each UAV in the swarm moves left (L), ahead (A)  or  right (R)  based  on the  pheromone  values  in  its respective  neighboring cells  and  the  next  value ($\rho_n$) in  the  first  return map  of  the  Rossler  attractor  (see Fig.~$1$ in \cite{CACOC2}). 

In CACOC\textsuperscript{2} model, the Boids flocking behavior \cite{boids}, including the collision avoidance, velocity matching and flock centering, is combined with CACOC to improve the network connectivity. Here, the flock centering forces the UAVs to maintain connectivity. The model uses two forces \cite{CACOC2} :  
\begin{itemize}
	\item $\hat{F}_{C}$ is a vector that gives a direction (L, R or A). 
	\item $\hat{F}_{flock}$  is a vector for the flock centering force computed with the average value of the last vector used for the neighboring UAVs. 
\end{itemize}
The normalized sum of the these two force vectors gives a vector $\hat{V}$ with a constant speed $v$ \cite{CACOC2} :

\begin{equation} \label{d_vector}
\hat{V} = v \cdot \frac{\hat{F}_{C} + f\cdot\hat{F}_{flock}}{\lVert\hat{F}_{C} + f\cdot\hat{F}_{flock}\rVert^2}
\end{equation}

In \eqref{d_vector}, $f$ represents the influence of flocking force, which determines the connectivity among the UAVs.

\section{Overview of Pheromone Mobility Model}\label{Overview_Pheromone_Mobility_section}

The pheromone mobility model uses repel digital pheromones to promote exploration and fast coverage of an area with no prior information \cite{rajan-b17}. Note that a digital pheromone has the same characteristics of a natural pheromone, such as deposition, evaporation and diffusion. Each UAV moves towards the cells with minimum repel pheromone value and deposits a repel pheromone of magnitude ‘1’ in the cells scanned along its trajectory. After a UAV deposits a pheromone in a cell $(x,y)$, it is progressively diffused to the surrounding cells, with a constant diffusion rate $\psi\in[0,1]$. This encourages UAVs to spread out and move toward the unvisited cells. The pheromone value of each cell also evaporates, decreasing its intensity over time by a constant rate $\lambda \in[0,1]$. If the map environment and target locations change with time, the evaporation of the deposited repel pheromones over time allows for UAVs to revisit already scanned cells of the map after a certain time gap. 

For simplicity, the UAVs are assumed to move in two-dimensional space to search a given area, which is divided in a grid of $C\textsuperscript{2}$ cells, where each cell is identified by its $(x,y)$  coordinates. Pheromones deposited by each UAV in the grid space are saved in a digital pheromone map. In a decentralized UAV network, the UAVs exchange their digital pheromone maps with their 1-hop neighbors by using the periodic 'hello messages'.

Mathematically the pheromone value $p_{(x,y)}$ in a cell $(x,y)$ at time $t$ is described as \cite{b2,b3,b7},
\begin{multline}
    p_{(x,y)}(t)=(1-\lambda)  \cdot [ (1-\psi) \cdot p_{(x,y)}(t-1)+\\
    \partial p_{(x,y)}(t-1,t) + \partial d_{(x,y)}(t-1,t) ]\label{pher_dynamic}
\end{multline}
where $(1-\psi) \cdot p_{(x,y)}(t-1)$ is the pheromone value remaining in cell $(x,y)$ after diffusion to the surrounding cells, $\partial p_{(x,y)}(t-1,t)$ is the new pheromone value deposited in the update interval $(t-1,t)$, and $\partial d_{(x,y)}(t-1,t)$ is the additional pheromone diffused to the current cell from its eight surrounding cells in the update interval $(t-1,t)$, which is described as,
\begin{equation}
\partial d_{(x,y)}(t-1,t)=\frac{\psi}{8} \cdot \sum_{a=-1}^1 \sum_{b=-1}^1 p_{(x+a,y+b)}(t-1) \label{pher_diffusion}
\end{equation}

\section{Connectivity-Aware Pheromone Model} \label{proposed_CAP}

\begin{figure}[htbp]
\centerline{\includegraphics[width=0.85\linewidth]{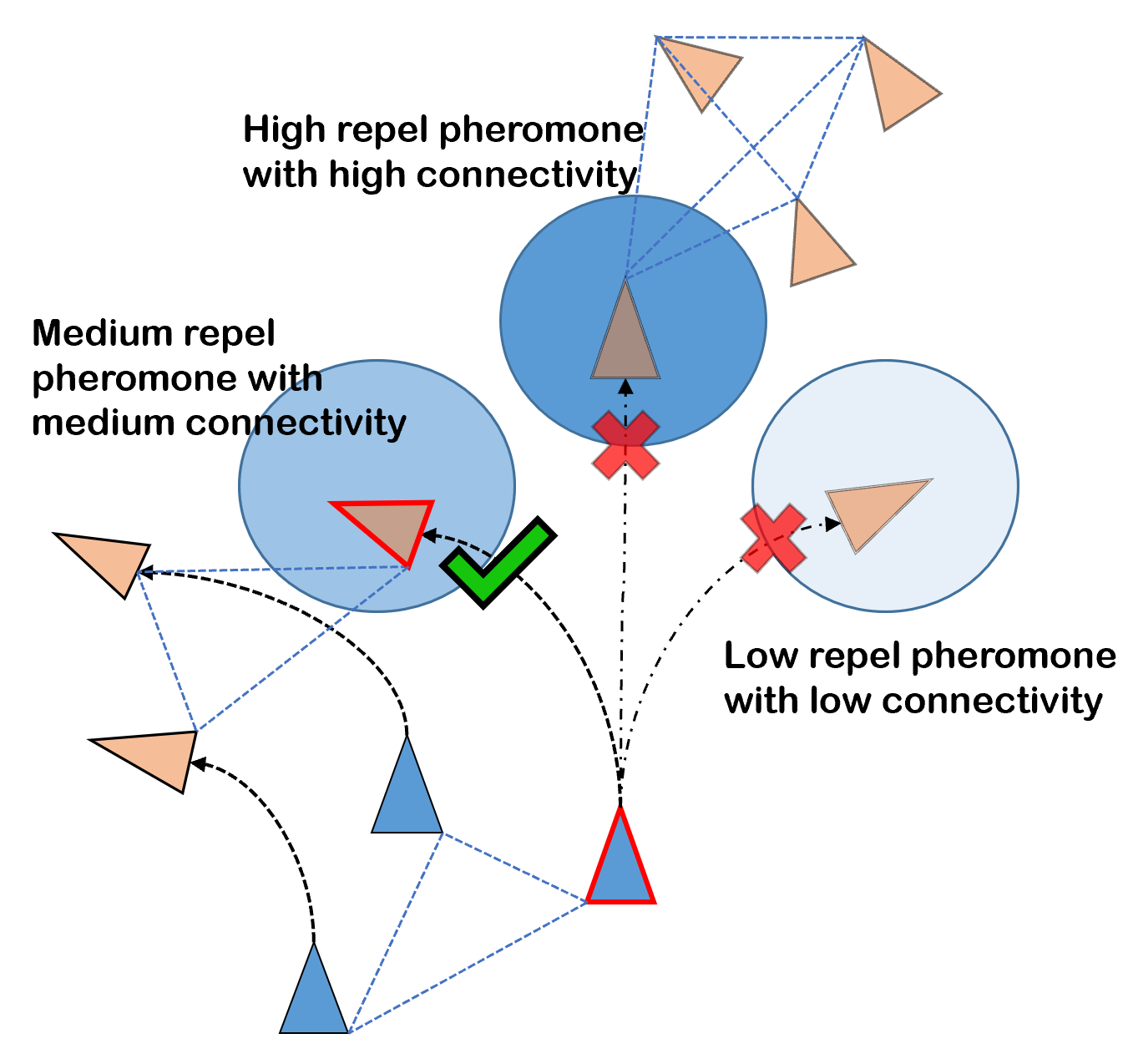}}
\caption{Illustration of next-waypoint selection based on repel pheromone intensity and connectivity of the UAV.}
\label{fig_conn-orient-pher-model}
\end{figure}

The pheromone mobility models achieve a fast coverage of the area by pushing the UAVs away from each other. However, this leads to poor connectivity among UAV nodes due to a limited transmission range of UAVs. Maintaining a strong connectivity among UAV nodes is essential in a decentralized autonomous UAV network to distribute pheromone information among its members and to achieve effective network coordination and communication between the UAVs. For example, if a UAV in Fig.~\ref{fig_conn-orient-pher-model} follows the repel pheromone gradient to a region of low connectivity, it is more likely to visit a region that has not been recently visited by another UAV. This would improve the coverage but the UAV may lose connectivity with other UAVs. Conversely, moving to a region with high repel pheromone where more UAVs are present would increase its connectivity, but the coverage performance will suffer.

In this section, we describe our proposed connectivity-aware pheromone mobility (CAP) model for decentralized autonomous UAV networks that maintains a strong connectivity as well as fast coverage. This model uses a weighted combination of the repel pheromone value and connectivity at the cells to guide the movement of UAVs.

\subsection{Next-Waypoint Cells}

The UAVs move from one cell to another, and the cell to which a UAV decides to move to is called the next-waypoint cell. After reaching a previously selected next-waypoint cell, a UAV selects its new next-waypoint cell. The distance (in terms of the number of cells) between the current and next waypoint cell is a function of the UAV speed and cell size.

\begin{figure}[htbp]
\begin{subfigure}{0.45\linewidth}
\includegraphics[width=\linewidth]{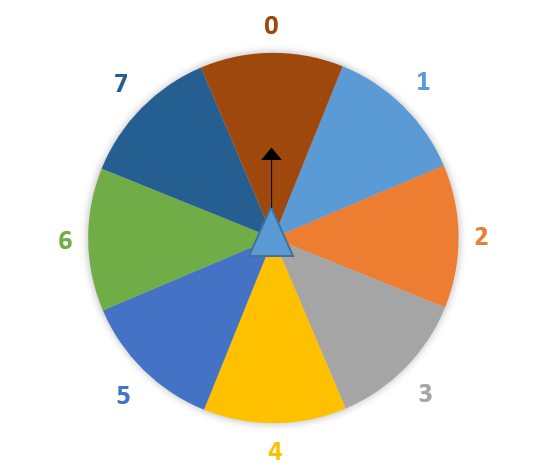}
\caption{}
\label{fig:headings}
\end{subfigure}
\begin{subfigure}{0.45\linewidth}
\includegraphics[width=\linewidth]{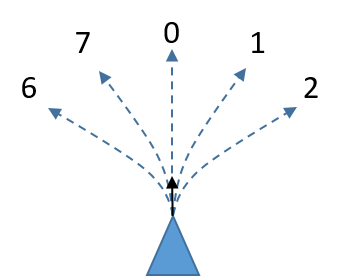}
\caption{}
\label{fig:nextwaypoints}
\end{subfigure}
\caption{(a) UAV heading discretized into 8 directions, (b) Selection of next-waypoint cells for a fixed-wing UAV.}
\label{fig:fig_heading_nextwaypoint}
\end{figure}

The heading of the UAVs (0 to 360 degrees) is discretized into 8 directions (see Fig.~\ref{fig:headings}), and the next-waypoint cell is chosen with reference to the current heading of a UAV such that it satisfies the flight trajectory constraints of a fixed-wing UAV, giving smooth turn trajectories. As shown in Fig.~\ref{fig:nextwaypoints}, the UAV with a current heading of ‘0’ selects one of the five possible next-waypoints cells (6, 7, 0, 1, and 2).

\subsection{``Look-Ahead'' Pheromone Value}

In pheromone mobility model, a UAV moves towards a cell with the minimum repel pheromone value. Instead, our scheme uses the \textbf{`look-ahead pheromone'} value, $P^{'}$, of the next-waypoint cells. The $P^{'}$ value of a next-waypoint cell $(x,y)$ is calculated as,
\begin{multline}
P^{'}_{(x,y)}=\frac{1}{12} \cdot (4 \cdot P_{(x,y)} +\\ (\sum_{(a, b\in[-1,1]\;\ except\ a=b=0)} P_{(x+a,y+b)})) \label{LA-pher}
\end{multline}
where $P_{(x,y)}$ is the pheromone value in cell $(x,y)$ and $P_{(x+a,y+b)}$ represents a pheromone value in eight 1-hop neighbors of cell $(x,y)$. 

By selecting the next-waypoint cell with minimum $P^{'}$ value, the UAV is more likely to visit cells that have not been visited before, thereby increasing its coverage performance.

\subsection{Distance-Weighted Degree-of-Connectivity}

The degree of connectivity of a UAV is a measure of the number of its 1-hop neighbors. For a UAV, the closer to the edge of the transmission boundary its 1-hop UAV is, the more likely it is to lose connectivity in near future. Therefore, we calculate the \textbf{distance-weighted connectivity} $(\gamma_{uv})$ between two UAVs ($u$ and $v$) as a function of their Euclidean distance $(d_{uv})$ and transmission range $(Tx)$. It is defined as,
\begin{equation} \label{deg}
\gamma_{uv} = 
\scalebox{.9}{$
\begin{cases} 
    1 & d_{uv} \leq (0.6\cdot Tx) \\
    2.5(1-\frac{d_{uv}}{Tx}) & (0.6\cdot Tx) < d_{uv} \leq Tx \\
    0 & d_{uv} > Tx
\end{cases}
$}
\end{equation}

$\gamma_{uv}$  is set to 1 when the distance between two UAVs is within 60\% of the transmission range, because the probability of the two UAV remaining connected is high. Value of $\gamma_{uv}$ decreases linearly when their distance ($d_{uv}$) exceeds 60\% of the transmission range.

Further, the \textbf{distance-weighted degree-of-connectivity} $K$ of a UAV $u$ is defined as the sum of its $\gamma_{uv}$ with all its 1-hop neighboring UAVs $\cal N$,
\begin{equation}
K = \sum_{v \in {\cal N}} \gamma_{uv}  \label{dst-deg}
\end{equation}

\subsection{Distributed Information Exchange using `Hello Messages'}

Hello message containing each UAV’s updated local information is propagated to its 1-hop neighbors. The pheromone and connectivity information of a UAV’s neighbors is used to select its next-waypoint cells and coordinate with its neighbors. In our scheme, each UAV exchanges the `Hello' messages with its 1-hop neighbors every 2 seconds, which consist of the UAV Id, its current location, next waypoint cell, and local pheromone map (pheromone value in the 5 x 5 cells centered at the UAV’s current cell).

\subsection{Selecting Next-Waypoint based on Look-Ahead Pheromone Value and Distance-Weighted Degree-of-Connectivity}\label{Pi_Ki_alpha_Section}

A UAV selects the next-waypoint cell $i$ with the maximum $W_i$ value among its 5 possible next-waypoint cells. Here, $W_i$ is defined as,

\begin{equation} \label{Pi_eq}
W_i = \alpha_i (1-P^{'}_i)
\end{equation}
where $P^{'}_i$ is the current `look-ahead pheromone' value at $i$ and $\alpha_i$ is defined as,
 
 \begin{equation} \label{alpha_function}
\alpha_i = 
\scalebox{1.0}{$
\begin{cases}
    \frac{K_i}{\beta} & K_i < \beta\\
    1  & K_i \geq \beta
\end{cases}
$}
\end{equation}

Here, $K_i$ is the estimated distance-weighted degree-of-connectivity of a UAV at the next waypoint cell $i$ (see \eqref{dst-deg})
and is calculated by using the heading information of its 1-hop neighbors received in the most recent hello messages. Varying the value of $\beta$ in \eqref{alpha_function} allows for tuning the connectivity and coverage performance of the proposed CAP model. We have varied $\beta$ from 0.5 to 4. Selecting $\beta$ as 4 gives a model with high connectivity performance that requires a longer coverage time, whereas selecting $\beta$ as 0.5 gives a model with high coverage performance and low connectivity. Here UAVs with $K$ $\geq$ 4 are considered to be sufficiently well connected with their neighbors. The next-waypoint selection process of a UAV is described in Pseudocode \ref{pseudo:cap}.

\begin{algorithm}
	\SetAlgorithmName{Pseudocode}{}
	\LinesNumbered 
	\setcounter{AlgoLine}{0}
	\begin{small}
		\uIf{UAV reaches next-waypoint cell}{
			// Deposit repel pheromone\\
			Add repel pheromone value = 1 for the current cell in its digital pheromone map\;
			// Select new a next-waypoint cell (i)\\
			Calculate the `look-ahead pheromone’$(P^{'}_i)$ value for each of the five possible next-waypoint cells\;
			Calculate the estimated distance-weighted degree-of-connectivity$(K_i)$ of the UAV at the five possible next-waypoint cells based on its neighbors next-waypoints\;
			Calculate the $W_i$ value for each of the five possible next-waypoint cells using \eqref{Pi_eq}\;
			Select cell $i$ with $maximum\ (W_i)$ value among the five possible next-waypoint cells as the UAVs next-waypoint cell\;}
		\Else{// Follow smooth-trajectory towards the selected next-waypoint cell}
	\end{small}
	\caption{UAV Next-Waypoint Cell Selection}
	\label{pseudo:cap}
\end{algorithm}

\section{Simulation Results and Discussion}\label{SimultionResults}

The performance of our proposed CAP model is compared against the repel pheromone and CACOC\textsuperscript{2} models. As explained in Section~\ref{CACOC2_section}, the CACOC\textsuperscript{2} model uses pheromone together with chaotic dynamics for improving coverage, and a flocking algorithm to maintain network connectivity. In contrast, our CAP model uses the pheromone to improve coverage and the local connectivity at the next-waypoint cells to maintain the overall network connectivity. 

\begin{table}
	\small
	\centering 
	\caption{Simulation Parameters}
	\label{sim_table}
	\begin{tabular}{l l}
	\toprule
		Parameters & Values  \\
	\midrule
		Number of Runs  &30 \\
		Simulation Time &8000 s \\
		Map Area 		&6 km $\times$ 6 km \\
		Cell Resolution & 100 m $\times$ 100 m \\
		Sensor Coverage Area &100 m $\times$ 100 m (1 cell) \\
		Transmission Range 	&1 km	\\
		Number of UAVs		&20, 30, 40\\
		UAV speed 	&20 m/s, 40 m/s	\\
		UAVs Start Postions &Mid bottom of map\\
		Evaporation Rate &0.006\\
		Diffusion Rate &0.006\\
	\bottomrule
	\end{tabular}\\
\end{table}

The UAV network simulation is implemented in Python3. Table~\ref{sim_table} shows the simulation parameters. For simplicity, the UAVs are assumed to be point masses, and their mobility is limited to the X-Y plane flying at a constant altitude. UAVs perform collision avoidance through trajectory modifications. A UAV scans the cell in which it currently resides and deposits a repel pheromone of magnitude $1$ in the scanned cell.

In the CAP model, a UAV selects its next waypoint every 10 s and 5 s for UAV speeds of 40 m/s and 20 m/s, respectively. For the CACOC\textsuperscript{2} model \cite{CACOC2}, we use  the 10 s  and 5 s update interval for 40 m/s and 20 m/s, respectively, to ensure a fair comparison with our proposed model.

\subsection{Performance Metrics}
The coverage performance of UAV network is measured by:
\begin{itemize}
\item \textit{Coverage Time (Tc)}: Average time taken to scan 90\% of cells in the map. 
\item \textit{Coverage Fairness (F)}: Represents how equally all the cells of the map are visited over a given time period, as measured by the Jain’s fairness index \cite{Jains_Index},
\begin{equation} \label{fairness_eq}
    F=\frac{(\sum_i x_i)^2}{n\sum_i x_i^2}
\end{equation}
where $x_i$ is the number of scans of cell $i$, and $n$ is the total number of cells in the map.
\end{itemize}

The connectivity performance is measured by:
\begin{itemize}
\item \textit{Number of Connected Components (NCC)}: Average number of disjoint components in the UAV network, sampled every 10 s. NCC measures how disconnected the UAV network is and its ideal value is 1.
\item \textit{Average Network Connectivity (ANC)}: Average number of links (or connected neighbors) each UAV maintains in the network, sampled every 10 s. 
\end{itemize}

\subsection{Results and Discussion}

\begin{figure*}
\begin{subfigure}{0.33\textwidth}
\includegraphics[width=\textwidth]{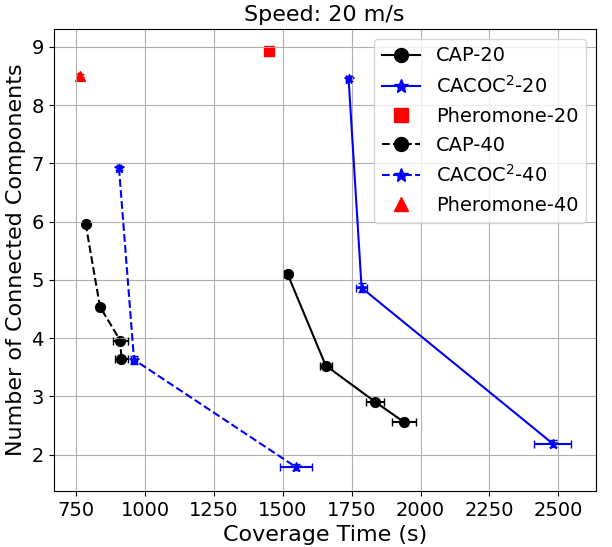} 
\caption{NCC}
\label{fig:ncc20-2040}
\end{subfigure}
\begin{subfigure}{0.33\textwidth}
\includegraphics[width=\textwidth]{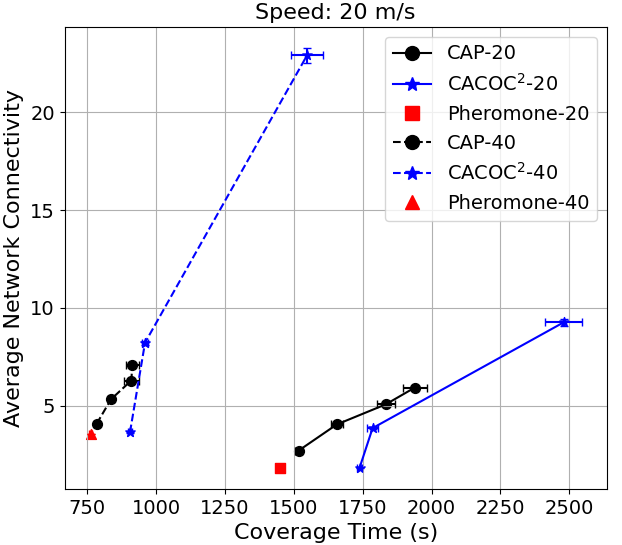}
\caption{ANC}
\label{fig:anc20-2040}
\end{subfigure}
\begin{subfigure}{0.33\textwidth}
\includegraphics[width=\textwidth]{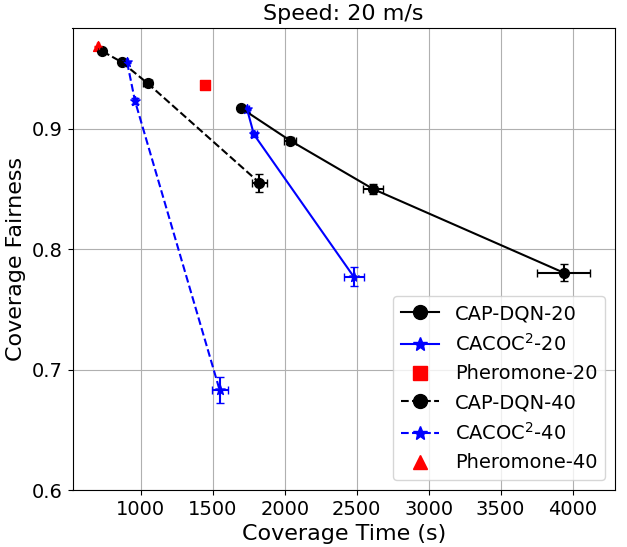}
\caption{Coverage Fairness}
\label{fig:f20-2040}
\end{subfigure}
\caption{Performance Curves for 20 and 40 UAVs at 20 m/s.} 

\label{fig:trade-off-s20n2040}
\end{figure*}

\begin{figure*}
\begin{subfigure}{0.33\textwidth}
\includegraphics[width=\textwidth]{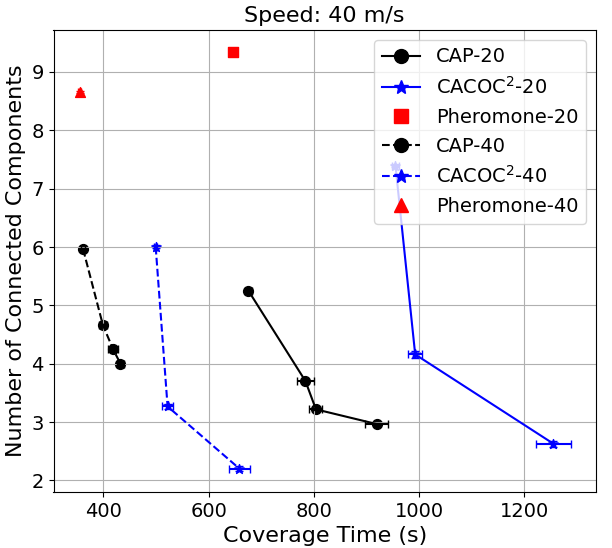} 
\caption{NCC}
\label{fig:ncc40-2040}
\end{subfigure}
\begin{subfigure}{0.33\textwidth}
\includegraphics[width=\textwidth]{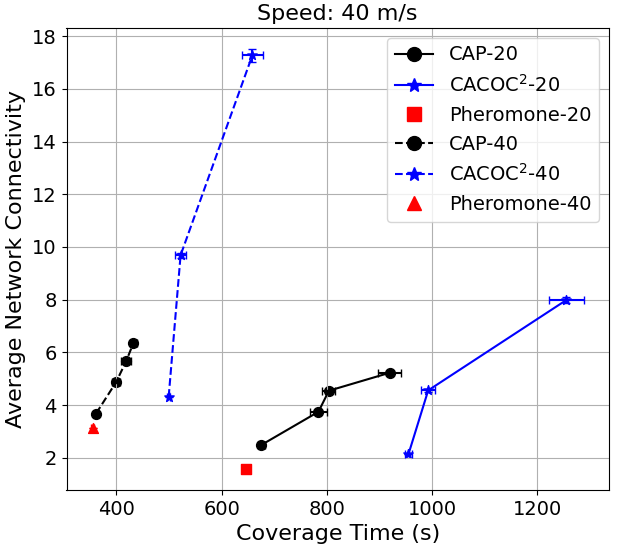}
\caption{ANC}
\label{fig:anc40-2040}
\end{subfigure}
\begin{subfigure}{0.33\textwidth}
\includegraphics[width=\textwidth]{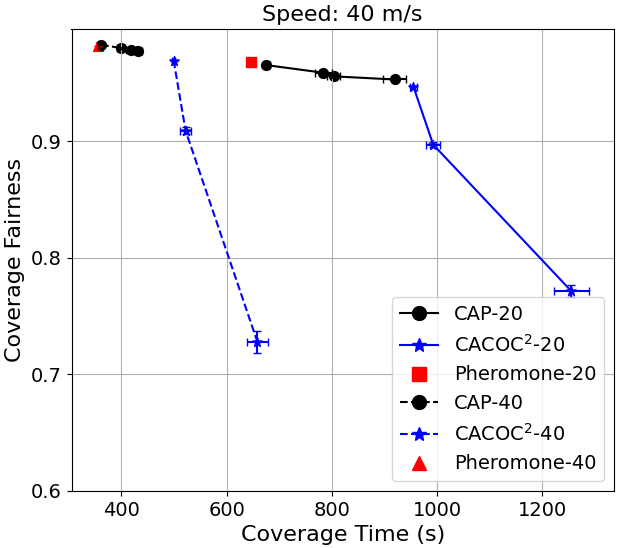}
\caption{Coverage Fairness}
\label{fig:f40-2040}
\end{subfigure}
\caption{Performance Curves for 20 and 40 UAVs at 40 m/s. }
\label{fig:trade-off-s40n2040}
\end{figure*}

We evaluate the coverage time (Tc), connectivity (NCC and ANC) and coverage fairness (F) performance for CAP, CACOC\textsuperscript{2} and the pheromone models for UAV densities of 20 and 40 UAVs, at speeds of 20 m/s and 40 m/s. The results are averaged over 30 simulation runs. Figures \ref{fig:trade-off-s20n2040} and \ref{fig:trade-off-s40n2040} show the performance curves (with error bars representing the standard error of the mean) of CAP (with $\beta = 0.5, ~2, ~3, ~4$) and CACOC\textsuperscript{2} (with $f = ~0.3, ~0.6, ~0.9$) models.
In the plots, the CAP-20, CACOC\textsuperscript{2}-20 and Pheromone-20 represent results for 20 UAVs, whereas CAP-40, CACOC\textsuperscript{2}-40 and Pheromone-40 represent results for 40 UAVs.

A low value of Tc and high value of F represents a better coverage performance, while a low NCC and high ANC indicate better connectivity. We consider that a model with $ANC \geq 4$ represents a strong connectivity, where the UAVs are sufficiently connected to their neighbours. In fact, a very high value of ANC may increase the co-channel interference and the probability of packet collisions during communication. 

The \textbf{NCC vs. Tc performance curves} for 20 and 40 UAVs at the node speed of 20 m/s are shown in Figure \ref{fig:ncc20-2040}. To ensure fast area coverage as well as good communication among the UAVs in the network, low NCC and high ANC values are desired at a low Tc value. For both UAV densities, the NCC vs. Tc curves of the CAP are closer to the lower left of the plot, demonstrating their superior performance compared to the CACOC\textsuperscript{2} model. For example, for 20 UAVs at 20 m/s (Fig.~\ref{fig:ncc20-2040}), at a Tc of around 1750s, the CAP and CACOC\textsuperscript{2} models achieve NCC value of around 3.2 and 7, respectively. For 40 UAVs (Fig.~\ref{fig:ncc20-2040}), the CACOC\textsuperscript{2} model achieves lower NCC values at much higher Tc values, but its coverage fairness performance is considerably worse (Fig.~\ref{fig:f20-2040}) due to the use of high flocking force. 

The \textbf{ANC vs. Tc performance curves} for 20 and 40 UAVs at 20 m/s are shown in Figure \ref{fig:anc20-2040}. The CAP model provides a higher ANC as compared to the CACOC\textsuperscript{2} model for a lower Tc value. For example, for 20 UAVs at 20 m/s (Fig.~\ref{fig:anc20-2040}), for a Tc of around 1750s, ANC values of around 5 and 3.8 were achieved by the CAP and and CACOC\textsuperscript{2} models, respectively. When using a higher flocking force (f), the CACOC\textsuperscript{2} model achieves higher ANC but its coverage performance degrades (higher Tc values). 

The \textbf{coverage fairness (F) vs. Tc performance curves} for 20 and 40 UAVs at 20 m/s are shown in Figure \ref{fig:f20-2040}. For both densities, the coverage fairness of our proposed CAP model is higher or comparable to CACOC\textsuperscript{2} model for lower Tc values. As connectivity increases (low NCC values) and the Tc values increase, the CAP model provides much better coverage fairness compared to the CACOC\textsuperscript{2} model. 
For a higher $f$ value, the coverage fairness of CACOC\textsuperscript{2} model decreases as the UAVs stick together, thus causing them to visit some cells more frequently than others. Our proposed CAP model therefore achieves fast as well as fair area coverage.

The performance trade-off curves at 40 m/s for 20 and 40 UAV densities are shown in Figure \ref{fig:trade-off-s40n2040}. Our proposed CAP model achieves even more pronounced improvements over CACOC\textsuperscript{2} model. The coverage time of both models improves at higher node speed, but the improvement is much higher for the CAP model, which also maintains better connectivity as well as a more fair coverage.

The coverage time (Tc) decreases and ANC increases with an increase in UAV density for both the models at both node speeds. However, the proposed CAP model provides better coverage fairness and a reasonably high ANC values, especially at higher UAV densities. Although the CACOC\textsuperscript{2} model achieves much higher ANC values at higher UAV density, a very high node connectivity does not necessarily improve the network connectivity performance.  

In summary, the repel pheromone model provides the best coverage but has a very poor connectivity performance. In general, the CAP model provides a fast coverage, which is closest to the pheromone model, while achieving a much higher connectivity than the pheromone model. For both UAV densities, our proposed CAP model provides better connectivity (lower NCC  and higher ANC) and fairness for a given Tc compared to the CACOC\textsuperscript{2} model. At low Tc values (fast coverage), the CAP model performs exceptionally well compared to the CACOC\textsuperscript{2} model.

\section{Conclusion} \label{conclusion_section}

We considered a decentralized, multi-hop, UAV network consisting of low SWaP fixed-wing UAVs. The area coverage and network connectivity
requirements of UAV networks exhibit a fundamental trade-off. To facilitate a reliable communication among UAVs in an autonomous UAV network, we designed a low-complexity connectivity-aware pheromone mobility (CAP) model. In CAP model, the UAVs make mobility decisions using a combination of the pheromone values and their local distance-weighted connectivity. It achieved an efficient coverage of the area, while preserving network connectivity and coverage fairness, and outperformed the CACOC\textsubscript{2} models at different UAV densities and speeds. 

Thus, the CAP model facilitates efficient inter-UAV communication and coordination of autonomous UAV networks for search and rescue operations.

\section{ACKNOWLEDGMENT OF SUPPORT AND DISCLAIMER}
This material is based on research sponsored by Air Force Research Laboratory (AFRL), under agreement No.  FA8750-18-1-0023 and FA8750-20-1-1005. The views and conclusions contained herein are those of the authors and should not be interpreted as necessarily representing the official policies or endorsements, either expressed or implied, of AFRL or the U.S. Government.

\end{document}